\documentclass[12pt,preprint]{aastex}
\usepackage{epsfig}

\newcommand{\arcm}{{$^\prime\,$}}
\newcommand{\arcs}{{$^{\prime\prime}\,$}}    
\newcommand{\z}{{z^\prime}}    

\newcommand{\CXO}{{\it Chandra}}
\renewcommand{\~}{{$\sim$}}

\begin{document}

\title{Shear-Selected Clusters From the Deep Lens Survey III: Masses
  from Weak Lensing}
 
\shorttitle{Shear-selected Clusters From the DLS}

\author{Alexandra Abate\altaffilmark{1}, 
D. Wittman\altaffilmark{2},
V. E. Margoniner\altaffilmark{2},
S. L. Bridle\altaffilmark{3},
Perry Gee\altaffilmark{2},
J. Anthony Tyson\altaffilmark{2},
Ian P. Dell'Antonio\altaffilmark{4}
}

\altaffiltext{1}{Laboratoire de l'Acc\'{e}l\'{e}rateur Lin\'{e}aire,
  IN2P3-CNRS, Universit\'{e} de Paris-Sud, BP. 34, 91898 Orsay Cedex,
  France; abate@lal.in2p3.fr} 
\altaffiltext{2}{Physics Department, University of California, Davis,
  CA 95616; dwittman@physics.ucdavis.edu} 
\altaffiltext{3}{Department of Physics,
  University College London, Gower Street, London WC1E 6BT}
\altaffiltext{4}{Physics
  Department, Brown University, Providence, RI 02912}

\begin{abstract}
We present weak lensing mass estimates of seven shear-selected galaxy
cluster candidates from the Deep Lens Survey.  The clusters were
previously identified as mass peaks in convergence maps of 8.6 deg$^2$
of $R$ band imaging, and followed up with X-ray and spectroscopic
confirmation, spanning a redshift range 0.19---0.68.  Most clusters
contained multiple X-ray peaks, yielding 17 total mass concentrations.
In this paper, we constrain the masses of these X-ray sources with
weak lensing, using photometric redshifts from the full set of $BVRz'$
imaging to properly weight background galaxies according to their
lensing distance ratios.  We fit both NFW and singular isothermal
sphere profiles, and find that the results are insensitive to the
assumed profile.  We also show that the results do not depend
significantly on the assumed prior on the position of the mass peak,
but that this may become an issue in future larger samples.  The
inferred velocity dispersions for the extended X-ray sources range
from 250---800 km s$^{-1}$, with the exception of one source for which
no lensing signal was found.  This work further establishes shear
selection as a viable technique for finding clusters, but also
highlights some unresolved issues such as
determination of the mass profile center without biasing the mass
estimate, and fully accounting for line-of-sight projections.  A
follow-up paper will examine the mass-X-ray scaling relations of these
clusters.
\end{abstract}



\section{Introduction}
Clusters of galaxies are the largest virialized structures in the
universe.  Tracing the distribution of these structures as a function
of cosmic time provides a measurement of the growth of structure and
is therefore a sensitive probe of dark energy \citep{HMH,b&w03}.  This
method of constraining dark energy requires accurate cluster mass
estimates as well as unbiased samples of clusters selected over a
broad range in mass and redshift (and with a well-defined selection
function).  There are two well-established selection
techniques---optical detection of the member galaxies
\citep{gal08} and X-ray detection of the hot
intracluster gas \citep{rosati02}---and two newer
techniques--- weak gravitational lensing \citep{wittman01,hs05} and
the Sunyaev-Zeld'dovich effect \cite[SZE, ][]{chr02}, in which the
cosmic microwave background is modified during its passage through the
large potential wells associated with the intracluster medium.

Each method has its strengths and weaknesses.  Optical detection is
capable of finding even poor clusters with relatively little
observational resources, but its cluster selection function is the
least well understood because the connection between mass (whose
clustering is predictable) and galaxies is not well understood.  X-ray
selection is the least affected by line-of-sight projections because
X-ray emissivity scales as the square of the local density, but by the
same token clumpiness within the cluster changes the observable at a
constant mass.  SZE has the advantage of being nearly
redshift-independent, but large SZE surveys are not yet available.
Lensing selection is based most directly on mass, but it is
observationally difficult because galaxies {\it behind} the cluster
must be imaged well enough to infer the shear induced by the cluster.
Whatever method is used to select the clusters, the next step is to
infer cluster masses.  Here the nonlensing methods depend on
assumptions about the dynamical state of the cluster, and lensing may
be useful for calibrating the masses of subsamples.  However, this is
observationally easier than determining the lensing masses of
individual clusters because subsamples may be stacked
\citep{johnston07,rozo08}.

In this paper we determine lensing masses of shear-selected
clusters, so hereafter we focus on the issues involved in shear
selection and lensing mass determination.  The weak lensing shear
signal induced by a cluster is a function of its mass profile and its
redshift relative to the background source galaxies being sheared, and
is independent of the nature of the matter and of its dynamical or
physical state.  However weak lensing mass estimates do suffer from
line-of-sight projections: structure along the line-of-sight could
artificially boost the lensing signal or cause false positive
detections.  In addition poor knowledge of the source redshift
distribution, as is often the case with only photometric redshift
estimates, limits the accuracy of the mass estimate.  We describe our
choices in dealing with these issues as they arise later in the paper.

This paper is organized as follows.  We describe the selection of and
previous work on these clusters in \S\ref{sec-sample}.  In
\S\ref{sec-data}, we summarize the data and its processing to the
point of shear and photometric redshift catalogues.  In
\S\ref{sec-fitting}, we describe the fitting procedure used to
estimate the masses.  We present the results in \S\ref{sec-results},
and summarize and discuss the implications in \S\ref{sec-discussion}.
Throughout this paper we assume a flat universe with $\Omega_m=0.3$
and Hubble constant $H_0=70$ km s$^{-1}$ Mpc$^{-1}$.

\section{Cluster Selection and Previous Work\label{sec-sample}}

The Deep Lens Survey \cite[DLS:][]{wittman02} is a deep 20 deg$^2$
$BVR\z$ survey conducted with the Mosaic imagers \citep{muller98} on
the CTIO and KPNO 4-m telescopes, with nominal exposure times of 18 ks
in $R$ and 12 ks in the remaining bands.  The clusters in this paper
were shear-selected from the 8.6 deg$^2$ which had at least 9 ks
coverage in $R$ band as of March 2002.  Here we briefly summarize their
selection and spectroscopic and X-ray followup.  We refer readers to
\cite{wittman06} for much more detail.

\cite{wittman06} ranked the shear peaks and then folded multiple peaks
within a 16\arcm\ \CXO\ ACIS-I field of view into candidate fields for
X-ray followup.  The top eight candidate fields were observed for 20
ks each with ACIS-I in \CXO\ AO4, with the exception of Abell 781,
which had archival \CXO\ data.  \cite{wittman06} also obtained Keck
and CTIO Hydra spectroscopy of likely cluster members, for redshift
confirmation and to identify any line-of-sight projections.  Redshifts
were found to range from 0.19 to 0.68.  Only one candidate (Candidate
5) was found to be a line-of-sight projection, on the basis of the
spectroscopy and the absence of any detected X-ray emission.  At the
same time, most other candidates were found to contain multiple
X-ray-emitting clumps, even where only one shear peak was present.  In
some cases, the additional extended X-ray sources correspond to
extensions of the convergence map contours away from the main peak.
See figures 8 to 12 and 14 to 17 in \cite{wittman06} where the multiband optical imaging of each candidate, with convergence maps and X-ray contours overlaid, were presented.%
The maps of \cite{wittman06} were not well suited to resolving
multiple peaks because of the heavy smoothing, as the kernel described
by \cite{wittman06} is approximately equal to a Gaussian with
$\sigma=6.25$\arcm.  In this paper, we use the shear data to fit a
mass model to each extended X-ray source.  This eliminates the need
for smoothing, so we expect this process to yield mass estimates even
where the original maps did not show a clear peak.  We also use
improved shear data as described below.

Two of the candidates have already been published in further detail.
\cite{wittman03} reported the spectroscopy and strong lensing arc of
Candidate 8 at $z=0.68$.  \cite{sehgal08} conducted a detailed X-ray
and shear-fitting analysis of Candidate 1 at $z=0.30$ (Abell 781; note
that the candidates are ranked by shear signal, so that Abell 781 was
the top candidate).  Abell 781 has multiple clumps which were fit
simultaneously so as not to bias the mass estimates for individual
clumps.  In this paper, we present simultaneous mass profile fits to
all seventeen X-ray-emitting clumps in all the candidate fields, as
well as an overall fit to each candidate field.  We repeat the fitting
of Abell 781 because \cite{sehgal08} used a completely different
code and a slightly inferior version of the data, and we show that the
two sets of results are consistent.


\section{Data}
\label{sec-data}

The DLS includes five widely separated 2$^\circ$ by 2$^\circ$ fields,
labeled F1 through F5.  F1 and F2 are in the northern hemisphere
and were observed with the Mosaic camera \citep{muller98} on the Kitt
Peak Mayall 4-m telescope, while the remaining fields are in the south
and were imaged with Mosaic II on the Cerro Tololo Blanco 4-m
telescope.  Each field is divided into a 3$\times$3 grid of
40\arcm\ by 40\arcm\ subfields.  The imaging data set for each
subfield consists of 20 dithered exposures in each of four bands
($BVR\z$), with each exposure lasting 900 s in $R$ and 600 s in the
other bands.  Observations were done in $R$ when the seeing was better
than 0.9\arcs\ FWHM, and galaxy shapes were measured in $R$ only.
\cite{wittman02} contains full details of the survey design.  As
mentioned above, the original candidate selection was done on the
subfields (totaling 8.6 deg$^2$) for which at least 9 ks (10
exposures) of $R$ band imaging had been completed by March 2002.  In
this paper, we use the full imaging data set, so the data are deeper,
and in some cases may extend to a larger distance from the candidate.

For each filter and subfield, the 20 exposures were combined into a
single stacked image 
using custom software as described in \cite{wittman06}.  %
The original images are 8192$\times$8192
pixels, with gaps between the eight CCDs and a pixel size which varies
over the field but averages to about 0.26\arcs.  The stacked images
are 10000$\times$10000 pixels with a uniform pixel scale of
0.257\arcs\ and with no gaps, cosmic rays, or satellite trails.  The
stacked point-spread function (PSF) size ranges from
0.8-0.9\arcsec\ in $R$ and from 0.9-1.2\arcsec\ in the other filters,
depending on field and subfield.  \cite{wittman06} contains many more
details on the stacking procedure.  A minor improvement over the
\cite{wittman06} and \cite{sehgal08} processing is the use of a
Lanczos-windowed sinc function for pixel interpolation, rather than a
truncated sinc function.\footnote{Lanczos interpolation is often used
  in popular packages such as SWarp, but the rationale is not widely
  known.  Sinc interpolation has the virtue of being {\it exact} for
  band-limited data \citep[e.g ][]{bracewell}, but it converges very
  slowly.  Simple truncation, as was done for the \cite{wittman06}
  images, leads to artifacts; one can see this by considering the
  effects in Fourier space of a top-hat window in pixel space.  The
  Lanczos-$n$ window smoothly truncates the sinc function at its $n$th
  zero by multiplying it by another sinc function whose first zero
  occurs there.  This also provides the useful property that the
  derivative of the interpolation kernel is zero at its edge.}

Cataloging and photometric redshift estimation proceeded much as in
\cite{sehgal08}.  We ran SExtractor \citep{b&a96} on the stacked
images in dual-image mode by detecting in $R$ and measuring in each of
$BVR\z$.  We determined shapes of galaxies using the ``VM'' method
described in \cite{heymans06}.  This is a partial implementation of
\cite{ellipto} which does not give higher weight to lower-ellipticity
sources, so it should not lose accuracy in high-shear regions as does
the full \cite{ellipto} method.  We apply a correction factor of
${1\over 0.89}$ to the VM shears, based on performance in the blind
analysis of simulations presented by \cite{heymans06} (corrected for
stellar contamination which was present in those simulations but not
in these data), and we fold a 10\% shear calibration systematic
uncertainty into the cluster mass uncertainties.

We used BPZ \citep{benitez00} to estimate photometric redshifts, using
the HDF prior and template tweaking as described in \cite{m&w08}
(based on the algorithm designed by \cite{ilbert06}).  When compared
against 328 galaxies with spectroscopic redshifts in the NASA/IPAC
Extragalactic Database ({\it not} used in the template tweaking), the
rms photometric redshift error per galaxy is $0.047( 1+z)$, with a
bias of $–0.017(1+z)$ and no catastrophic outliers over the redshift
range 0.02---0.70 \citep{sehgal08}.  However, these are relatively
bright galaxies, and we expect that performance is worse for the more
typical faint galaxies used in the shear fitting.  We address that
concern in \S\ref{sec:pz} below.

For each subfield, we matched the shape and photometric redshift
catalogs, eliminating galaxies which failed one or the other
procedure, and then applied masks around bright stars.  We also cut
sources with extremely large or exactly zero errors on the measured
ellipticity, indicating a problem with the ellipticity measurement;
saturated sources; and sources with a photometric redshift greater
than 1.6.  We imposed the latter cut because photometric redshift
performance is predicted to be poor after the 4000 Angstrom break is
shifted through the $\z$ filter, and we have no spectroscopy in that
redshift range to evaluate how it degrades.  Finally, we stitched
together the nine subfields in each field, checking that object
attributes in the overlap regions were consistent and that no objects
were duplicated.


\section{Fitting procedure}
\label{sec-fitting}

The observed reduced shear components, $g_1^o$ and $g_2^o$, are
calculated from the measured ellipticities of the source galaxies
$e_1$ and $e_2$ by dividing by the shear responsivity correction
$R_s$: $g_i^o=e_i/R_s$.  We fit a model for the mass distribution of
each cluster to the observed shear components, taking into account the
full three-dimensional position of each source galaxy (r.a., dec.,
photo-z).

We refer to each shear-selected cluster as a candidate and each X-ray
detection within the vicinity of a candidate position as a ``clump".
The X-ray positions give a better indication of a cluster center than
the shear peak position because the X-ray centroids are more precise
statistically.  \textit{Chandra} has 0.5\arcs angular resolution,
while the convergence maps have $>1$\arcm\ resolution.  We number the
candidates from 1 to 8 and label the X-ray clumps in their vicinity
with lowercase letters.  Where there are multiple shear peaks (which
have X-ray confirmation) within what could be considered a single
candidate, we label them with uppercase letters.  For each clump the
r.a. and dec. positions are given by the X-ray position, and the
redshift is given by a spectroscopic measurement, except for clumps
2b, 2c and 7c.  The photometric redshifts of these clumps are
consistent with the spectroscopic redshifts of the main clump in their
respective fields, so in these cases we use the main clump
spectroscopic redshift.  The information on the clusters according to
the labelling described above is presented in Table \ref{tb-clusters}.
Note that Table 2 of \cite{wittman06} contains a typographical error
in the position of Candidate 1; the correct position for the candidate
field is 09:20:50 +30:27:44 (J2000).

\begin{table}
\begin{center}
\caption{Top 8 shear-selected clusters from DLS. The clumps within
  each candidate are listed in order of decreasing X-ray
  flux. \label{tb-clusters}}
\begin{tabular}{|c|c|c|c|c|c|c|r|}
\hline
Cand. & \multicolumn{3}{|c|}{Shear peak info.} & \multicolumn{4}{|c|}{Clump info.} \\

\multicolumn{1}{|c|}{}& \multicolumn{1}{c}{Field} & \multicolumn{1}{c}{R.A.$^a$
} 
&  \multicolumn{1}{c|}{Dec.$^a$
} 
& \multicolumn{1}{c}{Clump} & \multicolumn{1}{c}{R.A.$^b$
} 
& \multicolumn{1}{c}{Dec.$^b$
} 
& \multicolumn{1}{c|}{z$^c$
} \\
\hline
1A& F2 & 09:20:27 & +30:30:44 & a& 9:20:26.4 & +30:29:39&0.302 (1) \\
1B&       &09:20:50  &+30:27:44  & b& 9:20:53.0 & +30:28:00&0.291 (1)\\
1C&       &09:21:11 & +30:27:39  & c& 9:21:10.3 & +30:27:52&0.427 (1)\\
                                             &&&&   d& 9:20:11.1 & +30:29:55&0.302 (1)\\
\hline
2 & F3 & 05:22:17 & -48:20:10 & a &05:22:15.6 & -48:18:17&0.296 (3) \\
                                             &&&& b &05:21:59.6 & -48:16:06&0.296 (2)\\
                                             &&&& c &05:21:47.6 & -48:21:24&0.296 (2)\\
                                             &&&& d &05:22:46.6 & -48:18:04&0.210 (5)\\  
\hline                                          
3 & F4 & 10:49:41 & -04:17:44 &a& 10:49:37.9 & -04:17:29&0.267 (3)\\
                                              &&& &b& 10:49:50.7 & -04:13:38&0.068 (4)\\
\hline
4 & F4 & 10:54:08 & -05:49:44  &a&10:54:14.8 & -05:48:50&0.190 (4)\\
\hline
5 & F5 & 14:02:12 & -10:28:14  &-&-&-&-\\
\hline
6 & F5 & 14:02:03 & -10:19:44  & a&14:01:59.7 & -10:23:02&0.427 (5)\\
\hline
7 & F2 & 09:16:00 & +29:31:34 &a& 09:15:51.8 & +29:36:37&0.530 (5) \\
                                                &&&&b&09:16:01.1 & +29:27:50&0.531 (5)\\
                                                &&&&c&09:15:54.4 & +29:33:16 &0.530 (2)\\
\hline
8 & F4 & 10:55:12 & -05:03:43  & a& 10:55:10.1 & -05:04:14&0.680 (6) \\
                                                &&&& b& 10:55:35.6 & -04:59:31&0.609 (5)\\
\hline
\hline
\end{tabular}
\end{center}
\leftskip=0.4cm \small{$^a$Position of shear peak in DLS.\\
$^b$Position of X-ray peak in Chandra follow-up.\\
$^c$Redshift sources: (1) Geller et al. 2005; (2) Wittman et al. 2006, photometric; (3) Wittman et al. 2006: CTIO 4-m/Hydra; (4) Colless et al. 2001: 2dF; (5) Wittman et al. 2006: Keck/LRIS; (6) Wittman et al. 2003}
\end{table}

\begin{figure}
\center
\includegraphics[width=10cm]{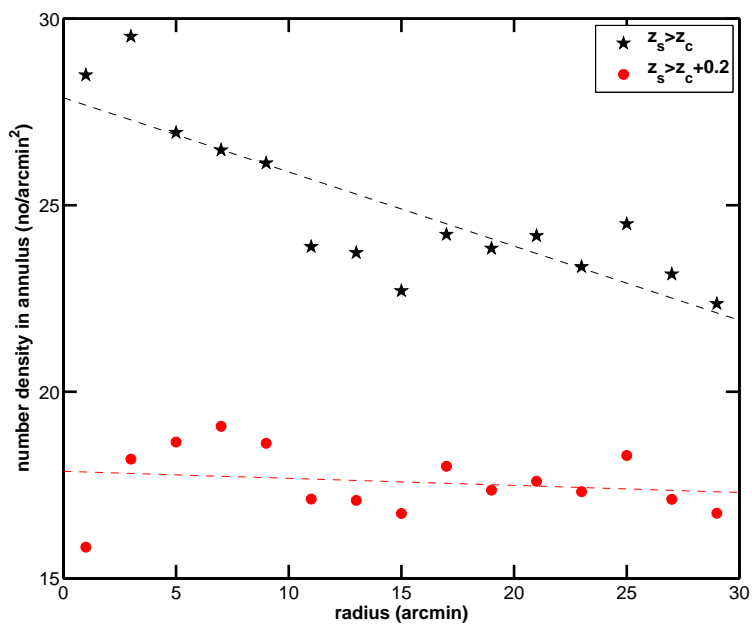} 
\caption[Cluster member contamination illustrated by the number
  density as a function of radius.]{Cluster member contamination
  illustrated by the number density as a function of radius for clump
  1a. The result is typical of the rest of the clusters.  The number
  density is plotted before (black stars) and after (red circles)
  applying a minimum photometric redshift cut on the source catalogue.
  The minimum photometric redshift cut alone is adequate in removing
  cluster member contamination. \label{fig:numdens}}
\end{figure}

In addition to the quality cuts mentioned in Section \ref{sec-data} we
remove any sources which have redshifts less than $z_c+0.2$, where
$z_c$ is the redshift of the cluster. This is to avoid cluster member
contamination in the sample of source galaxies.  Cluster members
scattering to higher redshifts would reduce the observed shear.  When
the fit applies to multiple clumps we take $z_c$ to be the redshift of
the highest redshift clump. The effect of the minimum photometric
redshift cut on cluster member contamination is shown by Figure
\ref{fig:numdens} for clump 1a; the same plot for the other clusters
yields similar results.  The number density as a function of radius is
plotted before and after applying a minimum photometric redshift cut
on the source catalogue.  When there is no minimum photometric
redshift (black stars) there is clear cluster member contamination at
radii less than 10 arcminutes. When the minimum photometric redshift
cut is applied (red circles) the number density is approximately flat
with radius as shown by the fitted red dotted line.  The gradient of
the red dotted line is not significantly different from zero given the
size of the scatter of the red circles.  This implies there is no
cluster member contamination, so no additional radius cut or
correction is required.

We use only sources that are within 30\arcm\ of a profile center,
partly for computational efficiency but mostly because sources at
large radii will contain very little information on the cluster mass
and be influenced by other nearby masses \cite[see Figure 8
  of][]{johnston07}.  Following \cite{hoek02} and \cite{cyp04} we
additionally remove sources within 1\arcm\ of a cluster center.  This
is because in the most inner regions the mass densities could be
closer to the critical value, leading to unmodeled strong lensing
effects.

For each model the best-fit parameters were obtained through
evaluation of the $\chi^2$ statistic, defined as
\begin{equation}
\label{eq:chisqe1e2}
\chi^2=\sum_i \left(\frac{g_1^p-g_1^o}{\sigma_{g_1}}\right)^2_i+\left(\frac{g_2^p-g_2^o}{\sigma_{g_2}}\right)^2_i
\end{equation}
where $g_1^o$ and $g_2^o$ are the measured shears. In
Eq.~\ref{eq:chisqe1e2} the $i$ denotes the different source galaxies.
The advantage of this approach is that it avoids data binning.  We
also use index $m$ to distinguish between the two ellipticity
components so $g^p_m$ is the predicted $g_1$ or $g_2$ component for
the source galaxy given a certain mass model. The error $\sigma_{g_m}$
is the error on the ellipticity component which comes from the shape
noise ($\sigma_{SN}$), the measurement noise ($\sigma_m$) and the
photometric redshift noise ($\sigma_{\gamma z}$, see Section
\ref{sec:pz} below) added in quadrature:
$\sigma_{g_m}^2=\sigma_{SN}^2+\sigma_m^2+\sigma_{\gamma z}^2$. A
separate measure of $\sigma_m$ for is used each galaxy.

The shape noise is estimated by finding the standard deviation of
$g_m$ values which have small measurement error
(i.e. $\sigma_m<0.002$, leaving about 700 galaxies to compute the
estimate).  Then the shape noise $\sigma_{SN}$ is the average value
estimated from $g_1$ and $g_2$ and we find $\sigma_{SN}\sim0.25$.

The predicted $g_1^p$ and $g_2^p$ are calculated from the reduced shear:
\begin{equation}
\label{eq:rs}
g=\frac{\gamma}{1-\kappa}
\end{equation}
where the form of $\gamma$ and $\kappa$ depends on the model (see Section \ref{sec:mm} below).
Then the prediction for each shear component $g_1^p$ and $g_2^p$ for a circularly symmetric distribution are:
\begin{eqnarray}
\label{eq:gnum}
g_1^p&=&-g\cos2\phi \nonumber\\
g_2^p&=&-g\sin2\phi
\end{eqnarray}
where $\phi$ is the angle subtended at the cluster center between the
positive x-axis and the line from the cluster center to the position
of the source galaxy.

\subsection{Mass Models}
\label{sec:mm}

We consider two different mass models, the singular isothermal sphere
(SIS) and the Navarro, Frenk and White (NFW) profiles.  The advantage
of the SIS profile is that it has a very simple form and only one free
parameter, the velocity dispersion $\sigma_v$. The NFW profile has two
free parameters, the concentration $c_c$ and $M_{200}$, the mass
within the radius at which the density is 200 times the critical
density in the universe.

The SIS profile is the density distribution found after solving the
Boltzmann equations describing gravitational equilibrium of an
isothermal gas.  For the SIS profile the convergence and the shear are
given by
\begin{equation}
\kappa=\gamma=\frac{1}{2}\frac{\theta_E}{|\theta|}
\end{equation}
where $\theta_E$ is the Einstein radius, which is related to the
one-dimensional velocity dispersion of the isothermal sphere by
\begin{equation}
\theta_E=\frac{4\pi \sigma_v^2}{c^2}\frac{D_{ds}}{D_s}
\end{equation}
$\sigma_v$ is the velocity dispersion (the mass proxy), $c$ the speed
of light, $D_{ds}$ is the angular diameter distance from the lensing
mass to the source galaxies, and $D_s$ is the angular diameter
distance from the observer to the source galaxies.  The shear from the
SIS profile diverges at small projected radii, but the enclosed mass
does not.  With the 1\arcm\ cut imposed for other reasons
(\S\ref{sec-fitting}), we do not attempt to fit these regions directly
in any case and therefore do not encounter numerical stability
problems.

The NFW profile is the distribution of dark matter predicted from
N-body simulations \citep{nfw}. The characteristic radius of a cluster
is defined as $r_s=r_{200}/c_c$, and the projected radius relative to
the center of the lens as $R$.  The analytic expression for the shear
as a function of $x\equiv R/r_s$ is given by \citep{w&b00}:
\begin{equation}
\label{eq:gammanfw}
\gamma(x)=
\begin{array}{ll}
\frac{r_s\delta_c\rho_{cr}}{\Sigma_{cr}}A(x) & \mbox{$\left(x<1\right)$} \\
&\\
\frac{r_s\delta_c\rho_{cr}}{\Sigma_{cr}}\left[\frac{10}{3}+4\ln\left(\frac{1}{2}\right)\right] &\mbox{$\left(x=1\right)$} \\
&\\
\frac{r_s\delta_c\rho_{cr}}{\Sigma_{cr}}B(x) &\mbox{$\left(x>1\right)$} 
\end{array}
\end{equation}
where the functions $A(x)$ and $B(x)$ depend only on the dimensionless
radius $x$ \cite[see][for their explicit form]{w&b00}, $\delta_c$ is a
characteristic overdensity for the halo, $\rho_{cr}$ is the critical
density of the universe at the redshift of the cluster and
$\Sigma_{cr}$ is the critical surface mass density,
\begin{equation}
\label{eq:sigcrit}
\Sigma_{cr}=\frac{c^2}{4\pi G}\frac{D_s}{D_d D_{ds}}
\end{equation}
where $D_d$ is the angular diameter distance from the observer to the
lensing mass.  The convergence $\kappa$ is the ratio of the local
value of the surface mass density to the critical surface mass
density, $\kappa=\Sigma(x)/\Sigma_{cr}$, where $\Sigma(x)$ is given
by \citep{w&b00}:

\begin{equation}
\label{eq:signfw}
\Sigma(x)=
\begin{array}{ll}
\frac{2 r_s\delta_c\rho_{cr}}{(x^2-1)}\left[1-\frac{2}{\sqrt{1-x^2}}\mbox{arc}\tanh\sqrt{\frac{1-x}{1+x}}\right] & \mbox{$\left(x<1\right)$} \\
&\\
\frac{2 r_s\delta_c\rho_{cr}}{3} &\mbox{$\left(x=1\right)$} \\
&\\
\frac{2 r_s\delta_c\rho_{cr}}{(x^2-1)}\left[1-\frac{2}{\sqrt{x^2-1}}\arctan \sqrt{\frac{x-1}{1+x}}\right] & \mbox{$\left(x>1\right)$} .\\
\end{array}
\end{equation}
The dependence on the halo mass enters Eq.~\ref{eq:signfw} and
Eq.~\ref{eq:gammanfw} mainly through the scale radius $r_s$, and the
dependence on the concentration mainly through $r_s$ and $\delta_c$.
There is a weak dependance on cosmology through $\rho_{cr}$,
$\Sigma_{cr}$ and $x$.

We perform the fitting in two ways:  
\begin{itemize}

\item[(i)] We fit one mass profile to each candidate. The high
  resolution of the X-ray data means that the X-ray position is a
  better estimator of the cluster center than the shear peak position.
  Therefore we vary the cluster center around the X-ray position; the
  cluster center is a free parameter.  The X-ray position will usually
  be decentered slightly from the observed mass peak because of noise
  peaks superimposed on the mass distribution \cite[see][and
    references within]{clowe06a}.  Therefore even using a well
  resolved shear peak position could cause a systematic overestimate
  of the mass.  If the offset is real, for example due to current or
  recent merging of clusters, then using the X-ray position would
  result in a systematic underestimate of the mass. At the same time,
  marginalizing over central positions over a large area biases the
  mass low. This is because fitting a profile positioned far from the
  true mass peak results in the prediction of a very low mass.  We
  have confirmed this effect using simulated data.

We therefore apply a prior on the cluster's X-ray/lensing offset, and
turn to the literature for guidance on the size of the prior.
\cite{smith05} examined this question observationally for ten massive
clusters with high-resolution lensing maps from \textit{Hubble Space
  Telescope} observations.  In this case, the X-ray/lensing offset is
much less affected by noise compared to the DLS clusters, which are
less massive and have lower-resolution lensing constraints due to
lower background source density. \cite{smith05} found offsets ranging
from 0---120 kpc, with an rms of 38 kpc.  \cite{koester07} examined 76
clusters identified in Sloan Digital Sky Survey photometric data and
found a typical offset of 57$h^{-1}$ kpc between the X-ray centroid
and the brightest cluster galaxy (BCG).  For $h=0.7$, the value
assumed throughout this paper, this is roughly double the separation
found by \cite{smith05}.  Assuming the BCG lies close to the bottom of
the gravitational potential, the \cite{koester07} result should be
more representative of our data because our clusters are of comparably
modest richness.

We therefore adopt a prior of the form
\begin{equation}
\label{eq:prior}
P=\mathcal{N} e^{-\frac{r^2}{2\sigma_r^2}} 
\end{equation}
where $r$ is the angular separation on the sky between the highest
flux X-ray peak position and the cluster center position, $\sigma_r$
is the angular separation of a 57$h^{-1}=81$ kpc distance at the
redshift of the cluster. We then marginalize over the cluster center
positions.
 
\item[(ii)] In the second approach, we simultaneously fit all the
  X-ray clumps in the vicinity of a candidate, to eliminate any
  influence one cluster may have on the inferred mass of another
  cluster.  The shears from different clusters add linearly because
  they are small, so the predicted shear for the multi-profile fit is
  $g_m^p=g_{ma}^p+g_{mb}^p+g_{mc}^p+...$, summing up all the predicted
  reduced shears from each cluster.  we use the exact X-ray peak
  positions as the profile centers in this fit.  The $g_{ma}^p$ etc
  are calculated as before (Eqs. \ref{eq:rs} and \ref{eq:gnum}) but
  the values of e.g. the distance ratios, $\phi$ and $|\theta|$ (used
  to calculate $g)$ will depend upon the clump in question.
\end{itemize}

\subsection{Photometric Redshift Errors}
\label{sec:pz}
Each of the source galaxies has a photometric redshift measurement and
enters into the fitting separately, and each photometric redshift has
an associated uncertainty that enters into the total uncertainty on
the mass model.  Here we calculate the relative magnitudes of the
uncertainties from the photometric redshifts and from each galaxy's
shear measurement.  For this estimate of the relevance of photometric
redshift errors only, we make the approximation that the observed
(reduced) shear is approximately equal to the shear $\gamma$,
\begin{equation}
g=\frac{\gamma}{1-\kappa}\simeq \gamma 
\end{equation}
where $\kappa$ is the convergence. The shear can be written as
\begin{equation}
\gamma(\theta) = \int  \cal{D}(\theta-\theta') \kappa(\theta')\textit{d}^{\rm2}\theta' ,
\end{equation}
\citep[see ][]{bridle98}, where $\cal{D}$ is the lensing kernel, and
$\kappa$ is the convergence defined in Section \ref{sec:mm}.  Using
the definition of $\kappa$ and Eq.~\ref{eq:sigcrit} one can write
\begin{equation}
\label{eq:gamdr}
\gamma(\theta)=\frac{4G}{c^2}\frac{D_d D_{ds}}{D_s}\int
\Sigma(\theta') \cal{D}(\theta-\theta')\textit{d}^{\rm2}\theta'
\end{equation}
or
\begin{equation}
\label{eq:gamdr2}
\int\Sigma(\theta') {\cal{D}}(\theta-\theta')\textit{d}^{\rm2}\theta'
=\gamma(\theta)\frac{c^2}{4G} \frac{D_s}{D_d D_{ds}}
\end{equation}
where the mass model is now isolated on the left-hand side and the
observables are isolated on the right-hand side. Uncertainties enter
the right-hand side on an equal basis through the estimate of the
shear $\gamma$ and through the distance ratio $D_{ds}/D_s$ which
depends on the source photometric redshift (the distance to the lens
$D_d$ is fixed through spectroscopic redshifts).  In our data, the
distance ratio uncertainty associated with a typical source galaxy's
photometric redshift measurement is about 30\% (expressed as a
percentage of its distance ratio).  The uncertainty in the shear
estimate from {\it any} source galaxy is much larger, $>$500\%,
because a typical shear is \~0.05 and shape noise---the uncertainty
floor set by the random orientations of source galaxies---is
\~0.25. Therefore the per-galaxy distance ratio uncertainty of
$\sim30$\% is negligible.

Although we have shown that random photometric redshift errors are
negligible on a per-galaxy basis, systematic errors in the photometric
redshifts can still cause a bias on the measured cluster masses.  To
address this question, we performed simulations of photometric
redshift errors similar to those in \cite{m&w08}.  We ran multiple
realizations tailored to each cluster, starting with the actual DLS
catalog for the subfield containing that cluster.  For each
realization, random photometric zeropoint errors were chosen according
to the zeropoint uncertainties in each filter as derived from repeat
observations.  One of two priors for the distribution of redshift and
spectral type as a function of magnitude were chosen: either the HDF
prior from \cite{benitez00}, or the prior from \cite{ilbert06}.  In a
given realization, for each galaxy, the $R$ magnitude was used to
choose a random type and redshift following the prior; these were used
to generate synthetic colors and photometric noise and zeropoint
errors were added.  The photometric redshifts were then estimated
using BPZ \citep{benitez00}, using either of the two priors at random,
thus simulating the effect of uncertain knowledge of the prior.

For a given lens, the induced shear scales linearly with the distance
ratio ${D_{ds} \over D_{s}}$ as shown by Eq.~\ref{eq:gamdr}.  Given
the true and estimated redshift for each galaxy in a simulation, the
ratio of true to estimated distance ratios was computed, and then
averaged over the galaxies in the simulation.  This yielded a distance
ratio bias for that realization.  The uncertainty in distance ratio
bias was estimated by taking the rms of thirty realizations for each
cluster.  The typical bias was 5-10\%, with the sign that the true
distance ratio was smaller than estimated, which implies that the mass
of the cluster is larger than estimated for a given shear.  The
typical uncertainty in the bias was 1-2\%.

We then use the distance ratio bias to ``correct" the distance ratio
calculated from the source photometric redshifts and the redshift of
the cluster in question.  The uncertainty in this bias correction is
negligible compared to that from other sources.  A simple error
propagation from the uncertainty in the correction to the mass
estimator shows a $\sim$1\% effect on the mass estimate.  The actual
uncertainty must be larger becuase the simulation does not include
some real effects, most notably a realistic variety of galaxy spectral
energy distributions (SEDs).  However, even if the uncertainty in the
source redshift bias were doubled or tripled, the loosening of the
mass constraints would be trivial, because other uncertainties
dominate.  At the same time, the 5-10\% correction does capture the
systematic effects of our filter gaps and photometry noise.


\section{Results}
\label{sec-results}

For the NFW profile, we found that the lensing data provide very
little constraint on the concentration parameter $c$.  We therefore
set the parameter to $c=5$, which is a typical value for the masses in
question \citep{seljak00,dolag04}.  Setting the concentration to 6
\citep[a high but reasonable value from][]{dolag04} instead of 5
changes the mass estimates by less than 10\%, well within the
uncertainty on the mass.  When fitting the SIS profile, stability
problems due to the divergence in the density at small radii were not
an issue, due to the central 1\arcm\ cut described above.  To compare
the results from the SIS fit and the NFW fit we convert the SIS
$\sigma_v$ into $M_{200}$.  They are related by
\begin{equation}
\label{eq:m200}
M_{200}=\sqrt{\frac{48}{200\pi\rho_{cr}}}\left(\frac{\sigma_v^2}{2G}\right)^{\frac{3}{2}}
\end{equation}
where $\rho_{cr}$ is at the redshift of the cluster.

We refer the reader to the figures presented in \cite{wittman06}, which show the multiband optical imaging of each candidate, with the convergence maps and X-ray contours overlaid. 

\subsection{Single Profile Fits}

\begin{figure}
\center
\includegraphics[width=10cm]{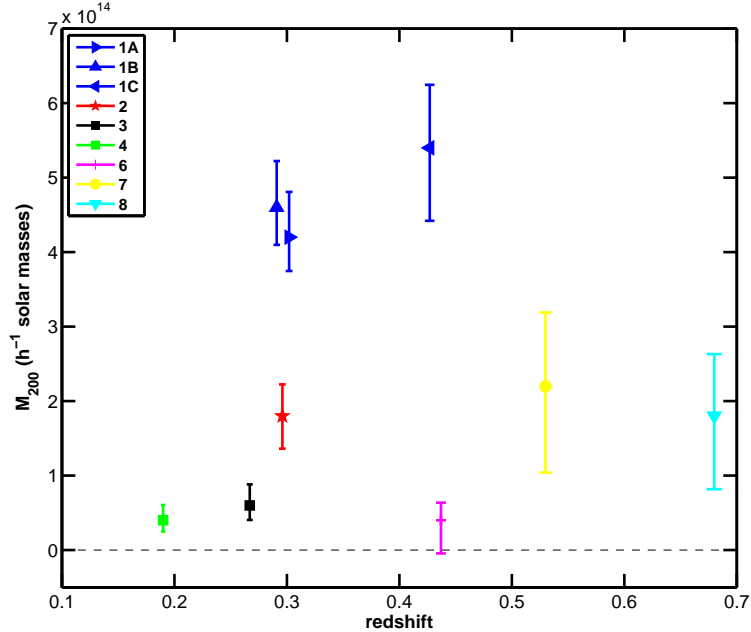}
\caption[Mass from the single NFW profile fit as a function of the
  cluster's redshift.]{Mass from the single NFW profile fit as a
  function of the cluster's redshift. See the legend to relate the
  data points to the candidates. \label{fig:singfit}}
\end{figure}

The results from the single profile fits, described in Section \ref{sec:mm} item (i), are presented in Table \ref{tb:sc} 
and Figure \ref{fig:singfit}.   
\begin{table}
\begin{center}
\begin{footnotesize}
\caption{Results from single profile fitting\label{tb:sc}}
\begin{tabular}{|c|c|c|c|c|c|c|c|}
\hline
Cluster & z &Centre offset&NFW Result
&  \multicolumn{2}{|c|}{SIS result} &Published result &No.\\

            &   & \tiny{(arcminutes)} &\tiny{$M_{200}$ (10$^{14}M_\odot/h$)} &\multicolumn{1}{c}{\tiny{$\sigma_v$ (km/s)}}&\multicolumn{1}{c|}{\tiny{$M_{200}$ (10$^{14}M_\odot/h$) }}& \tiny{$M_{200}$ (10$^{14}M_\odot/h$)} &sources \\
\hline
1A & 0.302 &0.52& 4.2$^{+0.6}_{-0.5}$  & 900$^{+37}_{-40}$  
      &  4.1$^{+0.5}_{-0.5}$ & 4.4$^{+1.6}_{-1.5}$$^a$ & 49411\\

1B &0.291 &0.34&4.6$^{+0.6}_{-0.5}$ &925$^{+42}_{-33} $
      & 4.5$^{+0.6}_{-0.5}$ & 3.7$^{+2.1}_{-1.5}$$^a$ &50456\\

1C &0.427 &0.40&5.4$^{+0.8}_{-1.0}$& 1025$^{+49}_{-60}$
      & 5.6$^{+0.8}_{-0.9}$& 4.3$^{+2.7}_{-2.4}$$^a$ &37648\\

2 & 0.296&0.31&1.8$^{+0.4}_{-0.4}$& 700$^{+44}_{-55}$
    &1.9$^{+0.4}_{-0.4}$ & - &45249\\

3 &0.267&0.43&0.6$^{+0.3}_{-0.2}$ & 400$^{+78}_{-109}$
  &0.4$^{+0.2}_{-0.2}$ & - &42743\\

4 &0.190&0.48&0.4$^{+0.2}_{-0.2}$& 225$^{+85}_{-141}$
  &0.1$^{+0.1}_{-0.1}$& - &40991\\

6 & 0.427&0.49&0.4$^{+0.2}_{-0.4}$& 300$^{+115}_{-206}$
  &0.1$^{+0.3}_{-0.1}$ &- &28626\\

7 & 0.530&0.31&2.2$^{+1.0}_{-1.2}$   &   800$^{+120}_{-118}$
   &2.5$^{+1.3}_{-0.9}$    &- &16354\\

8 &0.680&0.25&1.8$^{+0.8}_{-1.0}$& 725$^{+121}_{-134}$  
   &1.7$^{+0.9}_{-0.7}$ &2.8$^+_-$1.0$^b$ &19207\\
\hline
\hline
\end{tabular}
\end{footnotesize}
\end{center}
\leftskip=0.7cm \scriptsize{$^a$Converted from $M_{500}$ value in \cite{sehgal08}.\\
$^b$Converted from $\sigma_v$ in \cite{wittman03}. Note: We found there
was an error in the shear calibration of \cite{wittman03}. With
proper calibration, the weak lensing mass estimates and statistical
uncertainties in that paper should be divided by a factor of 1.6.  The mass presented here has had this correction applied.\\}
\end{table}

Table \ref{tb:sc} shows the resulting masses with 1$\sigma$ errors
after applying the prior for the cluster profile center position given
in Eq.~\ref{eq:prior}.  
We also include the offset in arcminutes between the best-fit centre position and the X-ray centre position; the average offset is about 0.4 arcminutes.  We find from simple simulations that a mis-center of this magnitude could still bias the mass estimate by up to 5 per cent. Therefore not applying the prior (Eq.~\ref{eq:prior}) might have been significant for some of the clusters in our sample, given the size of their statistical errors.

The mass found from the SIS profile is similar
to the one found from the NFW profile; the two mass estimates agree to
well within the $1\sigma$ level. 
The mass found for candidate 8 is
consistent with the estimate from \cite{wittman03} once the shear calibration error in that paper has been corrected, see footnote b in Table \ref{tb:sc}.  
Note that the previously published mass estimates for
Abell 781 \citep{sehgal08} are not directly comparable to the new
results presented in Table~\ref{tb:sc} because \cite{sehgal08}
performed simultaneous fitting, not sequential fitting of each clump
as in Table~\ref{tb:sc}.  We therefore defer a comparison to
\S\ref{sec:multiprofile}, in which we also fit all clumps
simultaneously.

Figure \ref{fig:singfit} presents the results from the single profile
fits using the NFW profile.  For each candidate $M_{200}$ with its
uncertainty is plotted at the cluster redshift.  The uncertainty in
the measured mass generally increases with redshift due to the
reduction in the number of sources available for the
analysis. Candidate 6 has a mass which is consistent with zero, and
candidates 3 and 4 have masses which are nearly consistent with zero.
This is a curious result for candidates which were selected on the
basis of their shear.  There are at least four effects at work here:

\begin{itemize}
\item The uncertainties here include systematic and not just
  statistical effects.  Thus, a candidate could be {\it detected} at,
  say 3$\sigma$, but with systematic uncertainties its mass estimate
  may be closer to 2$\sigma$ above zero.

\item The fits here marginalize over a range of possible mass centers,
  whereas the original detections are precisely at the locations with
  the highest (smoothed) shear value.  Noise fluctuations play some
  role in influencing this location, but at the same time,
  marginalizing over a large area will artificially lower the mass
  estimate and increase the uncertainty; see \S\ref{sec:mm} for a
  discussion of these issues.  Although we believe we have found a
  reasonable method for balancing these effects, the end result must
  be that the cluster mass estimate presented here has lower
  signal-to-noise than the original detection.

\item The original selection was made in the absence of photometric
  redshift information, while the mass estimates here use that
  information.  Even in the absence of projections, this introduces
  competing effects: sources at $z_{\rm phot}>1.6$ were thrown away
  here, thus reducing signal-to-noise if those sources were truly at a
  reasonably high redshift.  At the same time, signal-to-noise will be
  increased here to the extent that irrelevant sources (at redshifts
  less than the cluster redshift) can be discarded.

\item The use of photometric redshifts may correctly lower the
  signal-to-noise of projections which are widely enough separated in
  redshift.  Even clusters which are real would have preferentially
  entered the sample if other groups of galaxies were projected near
  their line of sight.  If the additional groups are at lower
  redshift, the cluster mass estimate will drop once source redshift
  information becomes available.  If the additional groups are at
  higher redshift, it would be difficult to eliminate their effect
  with the simple redshift cut adopted here; they would have to be
  specifically identified and modeled out.  The lowest-mass candidate
  in this paper, Candidate 6, in particular appears to have a
  foreground group of galaxies within a projected radius of 2\arcm.
  Candidate 3 has a foreground group which is much more visible in the
  images of \cite{wittman06}, and spectroscopically confirmed at
  $z=0.069$, but it is at a much larger projected distance (5\arcm).
  A tomographic search making use of source redshift information
  presumably would not have ranked these candidates so highly.  We
  will present the results of a tomographic search in a future paper.
\end{itemize}

\begin{figure}
\center
\includegraphics[width=7.5cm]{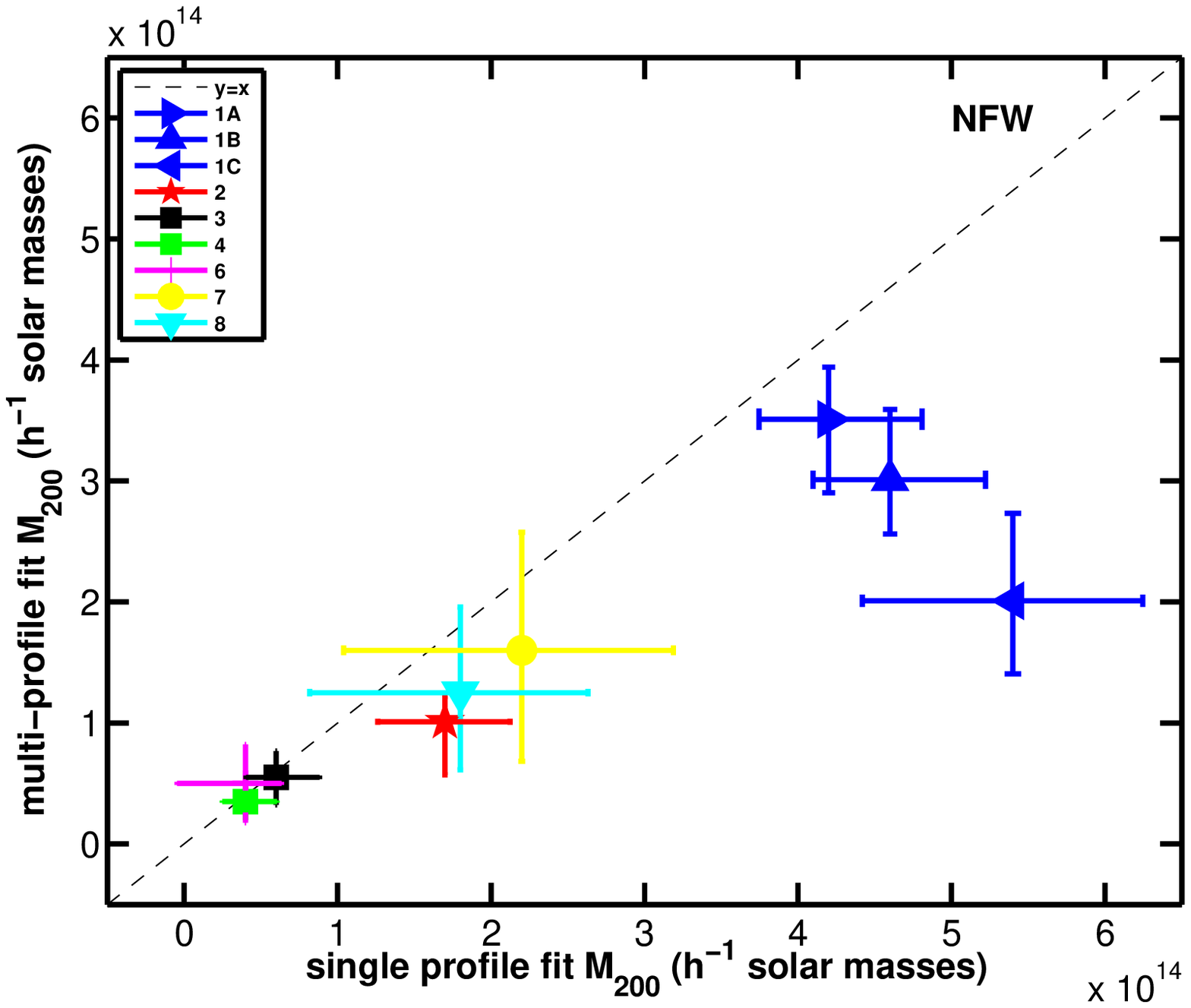}
\includegraphics[width=7.5cm]{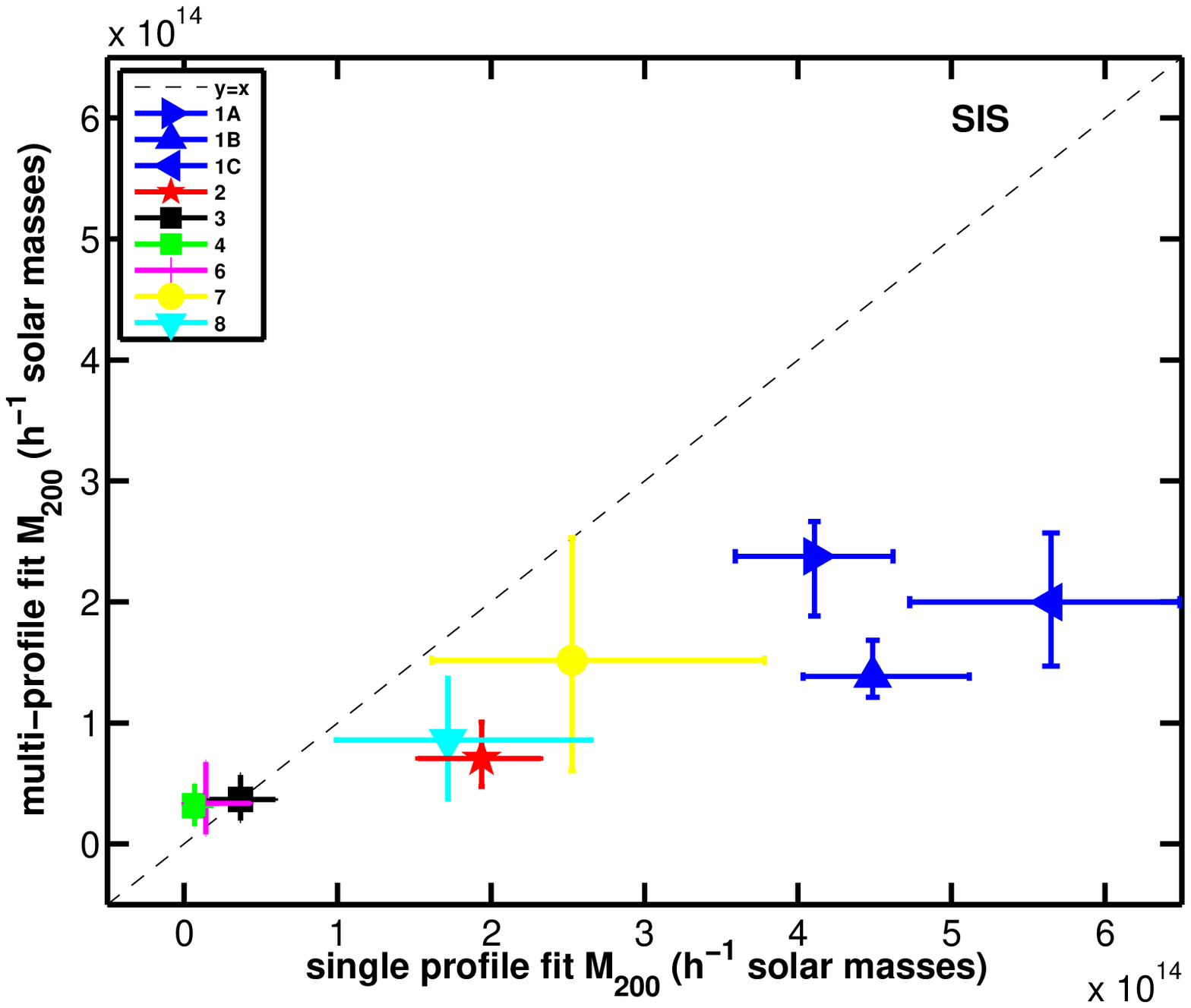}
\caption[Comparison of masses from the single profile fit and the
  multi-profile fit.]{Comparison of masses from the single profile fit
  and the multi-profile fit. The left hand panel shows the results
  from the NFW profile and the right hand panel shows the results from
  the SIS profile.  See the legend to relate the data points to the
  candidates.  Note that for clarity the candidate 2 result from the
  NFW profile fit (red/dark star point, left hand panel) has been
  offset to the left by 0.1x10$^{14}M_\odot/h$, which is much less
  than the size of its errorbar. \label{fig:smcomp}}
\end{figure}

\begin{figure}
\center
\includegraphics[width=10cm]{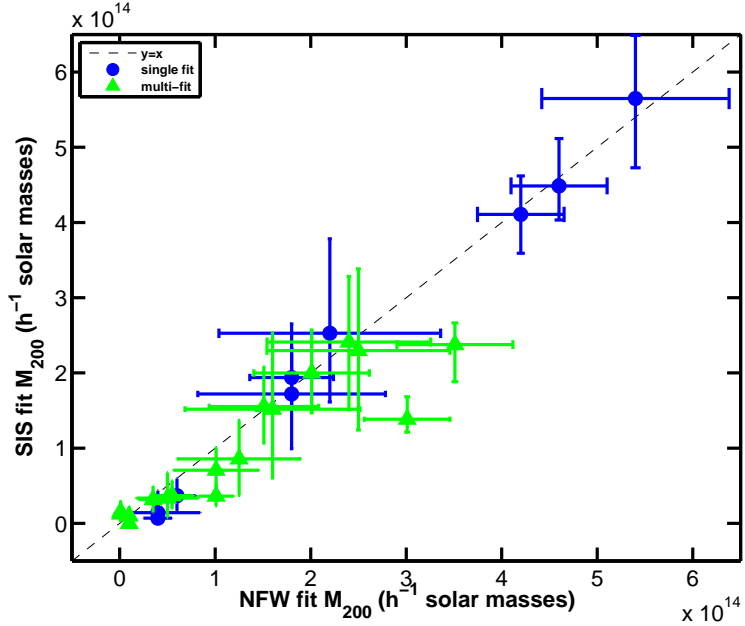}
\caption[Comparison of masses found from fitting the SIS profile and
  the NFW profile.]{Comparison of masses found from fitting the SIS
  profile and the NFW profile.  The green (light) triangle points are
  the results from the multi-profile fit.  The blue (dark) circle
  points are the results from the single profile fit.
\label{fig:compfit}}
\end{figure}

\begin{table}
\begin{center}
\begin{footnotesize}
\caption{Results from multi-profile fitting \label{tb:mc}}
\begin{tabular}{|c|c|c|c|r|r|c|c|}
\hline
Clump & Peak &z & NFW 
&  \multicolumn{2}{|c|}{SIS} &Published&No. \\

         &Value
         &     &\tiny{$M_{200}$ (10$^{14}M_\odot/h$)} &\multicolumn{1}{c}{\tiny{$\sigma_v$ (km/s)}}&\multicolumn{1}{c|}{\tiny{$M_{200}$ (10$^{14}M_\odot/h$)} }& \tiny{ $M_{200}$(10$^{14}M_\odot/h$)} &sources \\

\hline
1a & 0.0188 &0.302 & 3.5$^{+0.4}_{-0.6}$
&750$^{+30}_{-57}$&2.4$^{+0.3}_{-0.5}$
&4.4$^{+1.6}_{-1.5}$$^a$ &48228\\
1b &               &0.291 &3.0$^{+0.6}_{-0.4}$
&625$^{+43}_{-27}$&1.4$^{+0.3}_{-0.2}$
& 3.7$^{+2.1}_{-1.5}$$^a$&\\
1c &               &0.427 &2.0$^{+0.7}_{-0.6}$
&725$^{+64}_{-72}$&2.0$^{+0.6}_{-0.5}$
& 4.3$^{+2.7}_{-2.4}$$^a$& \\
1d &               &0.302 &0.0$^{+0.5}_{-0.0}$
&300$^{+80}_{-120}$&0.2$^{+0.1}_{-0.1}  $
& - &\\
\hline
2a & 0.0151 &0.296 &1.0$^{+0.2}_{-0.4}$
&500$^{+66}_{-66}$& 0.7$^{+0.3}_{-0.2}$       &-&51738\\
2b &               &0.296 &1.0$^{+0.6}_{-0.4}$
&400$^{+58}_{-61}$& 0.4$^{+0.2}_{-0.1}$    &-&\\
2c &               &0.296 &1.5$^{+0.5}_{-0.6}$
&650$^{+68}_{-79}$ &1.6$^{+0.5}_{-0.5}$    &-&\\
2d &               &0.210 &0.0$^{+0.4}_{-0.0}$
&275$^{+102}_{-126}$ &0.1$^{+0.2}_{-0.1}$  &-&\\
\hline
3a & 0.0136 &0.267 &0.6$^{+0.2}_{-0.2}$
& 400$^{+68}_{-86}$&0.4$^{+0.2}_{-0.2}$     &-&44680\\
3b &               &0.068 &0.1$^{+0.2}_{-0.2}$
& 250$^{+76}_{-127}$&0.1$^{+0.1}_{-0.1}$      &-&\\
\hline
4a &               &0.190 &0.4$^{+0.2}_{-0.2}$
& 375$^{+69}_{-97}$&0.3$^{+0.2}_{-0.2}$      &-&41085\\
\hline
6a &               &0.426 &0.5$^{+0.3}_{-0.3}$
& 400$^{+120}_{-188}$&0.3$^{+0.3}_{-0.3}$      &-&28632\\
\hline
7a & 0.0119 &0.530 &1.6$^{+1.0}_{-1.0}$
& 675$^{+146}_{-214}$&1.5$^{+1.0}_{-0.9}$      &-&19357\\
7b &               &0.531 &0.1$^{+0.3}_{-0.0}$
&0$^{+215}_{-0}$& 0.0$^{+0.2}_{-0.0}$   &-&\\
7c &               &0.530 &2.5$^{+1.2}_{-1.0}$
& 775$^{+117}_{-159}$ &2.3$^{+1.1}_{-1.1}$       &-&\\
\hline
8a & 0.0119 &0.680 &1.3$^{+0.7}_{-0.6}$
& 575$^{+113}_{-167}$ &0.9$^{+0.5}_{-0.5}$ &2.8$^+_-$1.0$^b$&20289\\
8b &               &0.609 &2.4$^{+0.9}_{-0.9}$
&  800$^{+91}_{-121}$&2.4$^{+0.9}_{-0.9}$            &-&\\
\hline
\hline
\end{tabular}
\end{footnotesize}
\end{center}
\leftskip=0.4cm \scriptsize{$^a$Converted from $M_{500}$ value in \cite{sehgal08}.\\
$^b$Converted from $\sigma_v$ in \cite{wittman03}, see footnote b in Table \ref{tb:sc}.\\}
\end{table}

\subsection{Simultaneous Multiple-Profile Fits\label{sec:multiprofile}}

The results from the multi-profile fit, described in Section
\ref{sec:mm} item (ii), are shown in Table \ref{tb:mc}.  Candidates 4
and 6 only had one X-ray peak so only one profile is fitted, but now
the profile center position is set as exactly the X-ray peak position.
The masses and 1$\sigma$ uncertainties in Table~\ref{tb:mc} are quoted
after marginalization over the masses of any other clumps within the
candidate field.  

Figure \ref{fig:smcomp} shows the relationship between the masses from
the single profile fit and multi-profile fit, with NFW results at left
and SIS results at right. Comparing the two panels, we see that
switching from single to multiple fits lowered the SIS mass estimates
more than it lowered the NFW mass estimates.  This suggests
empirically that the NFW profile is less susceptible to shear from the
neighboring clumps, a suggestion which is supported mathematically by
the greater steepness of the NFW profile at larger radii.

We now examine the effect of single vs. multiple fits on individual
clusters.  Candidates 1 and 2, with the most X-ray-emitting clumps
(four), are the two with significantly lower masses in the
multi-profile fit as compared to the single-profile fit.  This is
simply due to the shear of the neighboring clumps contributing
``mass'' to the single-profile fit.  Candidates 7 and 8, with three
and two X-ray-emitting clumps respectively, also have lower masses
with the simultaneous-fitting approach, but this effect is not
significant given their larger uncertainties.  For the remainder, the
mass estimates did not decrease with the simultaneous-fitting
approach; of these, only Candidate 3 had more than one X-ray-emitting
clump.  Collectively, the single-clump candidates saw their masses
{\it increase} slightly with the second approach. Although this is not
a very significant effect, it may follow from the one element of the
algorithm which changed, namely, that positions are fixed to the X-ray
positions rather than marginalized over.  The small difference between
the two results implies that the X-ray peak position was a good proxy
for the cluster center position, and that we did not marginalize over
too large an area.

The striking effect of clump multiplicity demonstrates that care must
be taken not to introduce a systematic error in the measured cluster
masses due to the presence of nearby shear peaks. It also implies that
there could be significant systematic errors from nearby undetected
masses.

We now compare our results to previously published mass estimates.
There is a dramatic change in Candidate 8.  Recall the
single-profile fit result of $1.7^{+0.9}_{-0.7} \times 10^{14} h^{-1}
M_\odot$ presented in Table~\ref{tb:sc} was higher than, but consistent with, the corrected \cite{wittman03} result of $2.8^+_-1.0 \times 10^{14} h^{-1}
M_\odot$.  When we simultaneously fit
the additional extended X-ray source in the field, the additional
source is assigned most of the mass, $2.4\pm0.9 \times 10^{14}
h^{-1}M_\odot$, and the main source is left with only $0.9\pm0.5\times
10^{14} h^{-1} M_\odot$.  The reason is unclear: the main source has a
larger X-ray flux by a factor of 1.4 despite being at a higher
redshift ($z=0.68$ vs. 0.61).  The dynamical and strong-lensing mass
estimates in \cite{wittman03} were also consistent with the higher
mass, and cannot be affected by the secondary cluster at lower
redshift.  It could be argued that those estimates do not strongly
rule out the lower mass estimate, given the small number of members
with spectroscopy (17) and the unknown redshift of the strongly lensed
arc.  Hence, a complete understanding of this system may require more
data, in the form of an arc redshift, {\it Hubble Space Telescope}
imaging for a higher-resolution weak lensing analysis, and/or more
spectroscopy. 

The second system with previously published mass estimates is Abell
781, for which \cite{sehgal08} simultaneously fit NFW profiles to
clumps 1a (which they called the ``Main'' cluster), 1b (which they
called the ``Middle'' cluster), and 1c (which they called the ``East''
cluster).\footnote{Note that \cite{sehgal08} considered cluster 1d to
  be merging with cluster 1a, and did not fit it separately.  We do
  fit it separately, which should lower the mass of cluster 1a
  somewhat.  \cite{sehgal08} also found a ``West'' cluster in XMM data
  covering a larger field.  We do not consider that cluster here, as
  its low mass and large projected distance make it unlikely to affect
  the results.}  Our simultaneously-fit NFW masses are consistent with
theirs.  Although Table \ref{tb:mc} gives the impression that the new
results are systematically somewhat lower, the total discrepancy of
the three clumps is only about one in terms of $\chi^2$, for three
degrees of freedom.  This is not significant.  The velocity
dispersions inferred from the multiple simultaneous SIS fits also
agree with the spectroscopic velocity dispersions measured by
\cite{geller05}.  When the shear fits are not simultaneous
(Table~\ref{tb:sc}), the inferred velocity dispersions are
significantly overstimated.

Finally, we compare the results between the two different profiles,
SIS and NFW, for all the measured masses (Figure \ref{fig:compfit}).
Only the masses of candidates 1a and 1b from the multiple profile fit
have more than a 1$\sigma$ discrepancy between the two profile types;
their SIS masses are lower than their NFW masses by about 2$\sigma$.
When the clumps are not fit simultaneously (blue points), masses
inferred from the two profiles are in excellent agreement.  Given the
greater susceptibility of the SIS to boosting by neighbors, we expect
a shift in this direction, but we would have expected the
non-simultaneous fits to lie {\it above} the unity-slope relation
rather than {\it on} it.  Given that these are the two highest-shear
clumps in the survey, this somewhat unexpected behavior may be
indicating that high-mass neighbors induce additional subtle effects
not accounted for by merely simultaneous fitting (for example,
magnification changing the source redshift distribution).

Figure \ref{fig:compfit} also suggests that the NFW profile tends to
assign more mass to the candidate; nearly all the multiple-profile
points lie below the line of unity slope.  For the single-profile fits, this effect seems to be nearly
cancelled by the tendency of the SIS to pull in more neighboring mass.
We investigated this observation with data simulated using an NFW profile.  We fitted the data with both SIS and NFW profiles and then compared the masses measured from both profiles.  We found instead that fitting the SIS profile caused a slight over-estimation of the true mass, but this was insignificant given the size of the statistical errors.  The over-estimation was due to the inner regions of the profile having more weight in the fitting, and in these regions the SIS density is less than the NFW density for the same mass. Therefore it is possible the reason why the NFW profile seems to assign more mass to the cluster than the SIS profile lies within the (unknown) exact details of the true density profile of clusters at these inner radii.  This is reasonable because the shape of the cluster density profile at the lowest radii is the most uncertain.  The NFW profile is predicted by cold dark matter only simulations, and would be affected at low radii by the baryonic cluster component, the existence of massive neutrinos or even the possibility that the dark matter has a non-zero cross-section.


\section{Summary and Discussion}
\label{sec-discussion}

We have measured the masses of a sample of shear-selected clusters
from the Deep Lens Survey by fitting NFW and SIS profiles to the full
dataset of shears and source redshifts.  For each type of profile, we
fit the data in two different ways: (i) we fit to each {\it shear peak
  separately}, marginalizing over likely locations of the mass profile
center; (ii) we fit to all the {\it extended X-ray sources
  simultaneously}, fixing the mass profile locations at those of the
X-ray peaks. We find that:
\begin{itemize}

\item Masses measured with either profile are consistent with each
  other on a per-cluster basis.  This is because the two profiles are
  similar to each other at the cluster radii probed by the weak
  lensing.  There are hints that the NFW profile tends to assign more mass on
  average, but a larger and higher-S/N sample would be required to
  confirm this. The same tendency was found by \cite{okabe09}, who 
  studied 22 clusters with weak lensing signal measured from 
  Subaru/Suprime-Cam imaging data.

\item Fitting multiple clumps simultaneously can yield a much lower
  mass estimate compared to fitting them separately.  This is because
  the single-profile fit conflates shear from nearby peaks with shear
  from the peak being fit.  The size of this effect was a factor of about 1.5 to
  2 for Abell 781 and for Candidate 2.

\item Uncertainty in the mass profile center had a relatively small
  effect here, but should not be overlooked.  For the candidates with
  a single extended X-ray source, the {\it only} difference between
  approaches (i) and (ii) was the marginalization over, vs. fixing of,
  the mass profile center, and the results did not change
  dramatically.  There was a shift in the sense that marginalizing
  over possible centers did yield lower mass estimates, but it was
  small compared to the typical per-cluster uncertainty here.  While
  adequate for the current purpose, with larger samples this may
  become a nontrivial systematic issue.  
  Especially since we find from simplified simulations that a small offset
  of just 0.4 arcminutes could cause a systematic bias on the mass if the
  precision of the statistical errors is greater than 5 per cent.  More 
  investigation into the typical BCG/X-ray and mass peak position offsets 
  is required to make sure this systematic is fully under control.

\item A few clusters have masses which are consistent or nearly
  consistent with zero.  Although several effects may be contributing,
  the most likely is that the original shear selection, having been
  done without regard to source redshifts, was vulnerable to
  line-of-sight projections in a way that the current fitting
  technique is not.  This problem may have been compounded by the
  heavy smoothing in the \cite{wittman06} convergence maps.  As shown
  by the spectroscopy and X-ray detections in \cite{wittman06}, these
  are real clusters, but they may be so poor that they passed the
  selection only because of nearby projections.  Tomographic selection
  should reduce this problem, but, according to the simulations of
  \cite{hs05}, cannot eliminate it.  Followup spectroscopy and/or
  X-ray observations can prove that a cluster exists at that location,
  but much more extensive (perhaps prohibitive) followup would be
  required to prove that that are no additional groups projected
  nearby that boosted the cluster into the sample.  An alternative
  would be to think of these systems simply as shear peaks rather than
  clusters, and use the formalism of \cite{wang09} or \cite{marian09}
  to constrain cosmology.
 
\end{itemize}

Comparing to previous mass estimates of some of these systems, we find
that Abell 781 is entirely consistent. However, Candidate 8 is more
puzzling than previously thought; the new fits indicate that the
secondary X-ray peak has more mass than the primary one.  More data
may be required to fully understand this system.

Simulations suggest that the ellipticity of clusters could be used as a cosmological probe, if sufficiently good statistics can be obtained \citep{HBB05,HBB06}.
Observations do suggest that cluster shapes, measured by the distribution of galaxies, X-rays, SZ effect and weak lensing, are usually not spherical.  There is also indication that the mean ellipticity evolves with redshift \citep{melott01,plionis02}, with clusters becoming less elliptical over time. This is expected because of hierarchical cluster formation and the reduced merger rate as clusters undergo a relaxation process. The systematic ellipticity of cluster dark matter halos was recently detected by stacking hundreds of clusters in SDSS \citep{EB09}. 

An extension to the current paper would be therefore, to allow the individual clumps to be fitted with an elliptical profile instead of an axi-symmetric one.  This may reduce the number of clumps required in the fit, or produce more realistic mass estimates for the individual clumps. However the ellipticity of clusters is difficult to measure for individual systems \cite[although see][]{cyp04} due to the high noise levels on individual shear measurements.  \cite{CK09} looked at the resulting biases on the cluster mass function from fitting a spherical model to both spherical and elliptical (triaxial) halos after stacking the lensing signals of many clusters binned by mass-correlated observables.  They fitted a circular model to the stack of elliptical halos and measured $M_{200}$ in a circular contour and found that this did not match with the $M_{200}$ values of the original individual halos, if this $M_{200}$ was calculated in elliptical isodensity contours. However they found that the stacked $M_{200}$ matched well to the $M_{200}$ of the individual halos, if the $M_{200}$ values of the original elliptical halos were defined using circular contours instead, where the density at different points around these circular contours was not constant.

We performed a simulation to test this effect on a per-cluster basis.  The clusters in \cite{CK09} were stacked without attention to aligning their axes, so the random orientations of many halos in each bin would artificially circularise the average of the stacked halos.  
In our simulations we calculate the difference between (i) the mass obtained from fitting a circular model and (ii) the mass within the circle containing $200\rho_{crit}$ of the best-fit elliptical model. We find these two numbers are the same to better than 1 per cent accuracy.
Therefore we conclude that there is negligible systematic bias in our results from fitting a spherical profile to an elliptical halo as long as $M_{200}$ in circular radii is the quantity of interest.

Many different cluster surveys will be taking place in the near
future.  They will aim to probe the growth of structure over time,
beginning a new genre of cosmological analysis.  Understanding the
sample selection and relating the cluster observables to the mass is a
large task.  Shear selection in principle eases this task because it
is independent of the nature of the matter and of its dynamical state,
but there are clearly several issues which must be resolved,
principally the observationally difficult task of accounting for
projections, but also including determination of the mass profile
center without biasing the mass estimate. In the future, large surveys
such as the Large-aperture Synoptic Survey Telescope (LSST) will find
on the order of 100,000 shear-selected clusters \citep{tyson03}.  With
such massive surveys systematics will become important.  The modest
samples produced by DLS and other current surveys
\citep{miy07,g&s07,dietrich07} will be the testing ground for
developing methods to successfully analyze these new large data sets.

\acknowledgments

We thank Jim Bosch and Eduardo Cypriano for useful conversations.  The
DLS has received generous support from Lucent Technologies and from
NSF grants AST 04-41072 and AST 01-34753 and NASA grant
NNG05GD32G. Observations were obtained at Cerro Tololo Inter-American
Observatory and Kitt Peak National Observatory, which are operated by
the Association of Universities for Research in Astronomy, Inc., under
cooperative agreement with the National Science Foundation.  AA
acknowledges support from a British Council grant and thanks the
University California, Davis for its hospitality during the completion
of this work.  SB acknowledges support from the Royal Society in the
form of a University Research Fellowship.

\end{document}